\documentclass[10pt,journal]{IEEEtran}
\usepackage{ifpdf}

\usepackage{cite}

\ifCLASSINFOpdf
   \usepackage[pdftex]{graphicx}
   \DeclareGraphicsExtensions{.pdf,.jpeg,.png}
\else
   \usepackage[dvips]{graphicx}
   \DeclareGraphicsExtensions{.eps}
\fi
\usepackage{amsmath}
\ifCLASSOPTIONcompsoc
  \usepackage[caption=false,font=normalsize,labelfont=sf,textfont=sf]{subfig}
\else
  \usepackage[caption=false,font=footnotesize]{subfig}
\fi

\usepackage{stfloats}
\usepackage{url}

\usepackage{multicol}
\usepackage{multirow}
\usepackage{amssymb}
\usepackage{colortbl}
\usepackage{tabularx}
\usepackage{lscape} 
\usepackage{pdflscape}
\usepackage{booktabs}
\definecolor{row1}{rgb}{0.95,0.95,0.95}
\definecolor{row2}{rgb}{0.85,0.85,0.85}

\begin{document}
%
\title{Building a SDN Enterprise WLAN Based on Virtual APs}
%
%
%

\author{Luis~Sequeira, Juan~Luis~de~la~Cruz, Jos\'{e} Ruiz-Mas, Jose Saldana, Juli\'{a}n Fernandez-Navajas and Jos\'{e} Almodovar
\thanks{L. Sequeira, J. Cruz, J. Ruiz-Mas, J. Saldana, J. Fernandez-Navajas are with CeNITEQ Group - Aragon Institute of Engineering Research (I3A), EINA University of Zaragoza, Spain. Email: \{sequeira, jlcruz, jruiz, jsaldana, navajas\}@unizar.es }
\thanks{J. Almodovar is with Networks Department, TNO, Netherlands. Email: jose.almodovarchico@tno.nl}
\thanks{This work has been partially financed by the EU H2020 Wi-5 (Grant Agreement no: 644262), TISFIBE (TIN2015-64770-R) Government of Spain, European Social Fund and Government of Aragon (research group T98).}
\vspace{-0.8cm}
}

\maketitle

\begin{abstract}
In this letter the development and testing of an open enterprise Wi-Fi solution based on virtual APs, managed by a central WLAN controller is presented. It allows seamless handovers between APs in different channels, maintaining the QoS of real-time services. The potential scalability issues associated to the beacon generation and channel assignment have been addressed. A battery of tests has been run in a real environment, and the results are reported in terms of packet loss and delay.
\end{abstract}

\vspace{-0.2cm}

%
\IEEEpeerreviewmaketitle

\section{Introduction}
%
%
%
%

\IEEEPARstart{D}{uring} the last few years, the telecom world has experienced an unexpected and unprecedented boom in the use of wireless devices (e.g. smartphones and tables) and the consequent deployments of extensive wireless networks in areas such as business centers, airports, malls, campuses or even entire cities. This situation has made clear the need for solutions including a set of coordinated Wi-Fi Access Points (APs from now), usually known as ``Enterprise Wi-Fi". Although commercial solutions exist, these are proprietary, closed and costly, which in most of the cases make them unfeasible for many organizations.

In this context, the scientific community is looking for proposals for inter AP coordination solutions enabling advanced features; e.g. load balancing, frequency planning or power control, while making use of low-cost hardware and open software. Some works have proposed the adaptation of certain abstractions and concepts from Software Defined Networks (SDNs), such as the concept of flow, for its use in wireless networks \cite{sdn1, sdn2}. However, important modifications and extensions are needed, since SDNs do not capture by themselves all the issues appearing in wireless scenarios (interference, mobility, channel selection, etc.). 

One of the first problems that arise when coordinating a wireless network is that the stations (STAs from now) have their own algorithms for selecting the AP. Therefore, each STA is free to select the AP to associate, purely on the basis of local decisions, not coordinated with the rest of the clients. This complicates client management, and results in problems; e.g. the ``sticky client", that never leaves the AP to which it was associated initially. Other problem is that the normal handoff may incur a delay up to several hundreds of milliseconds \cite{handoff1}. This fact may not pose a big problem for certain services, but it may constitute a severe limitation for real-time applications as VoIP or online games.

One way to solve these problems is the introduction of the \textit{Light Virtual Access Point} (LVAP) abstraction, first proposed in \cite{lvap1}. The idea is that a physical AP will use a different LVAP (which includes a specific MAC) for communicating with each STA. Therefore, the STA will only ``see" a single AP, even if it is actually moving between a set of them, thus avoiding the need for re-association.

It can be said that the LVAP ``travels" with the STA; i.e. it is dynamically assigned to a physical AP near the current location of the terminal. As long as the STA only ``sees" a single AP, it will not make any roaming decision, thus enabling the network to run a coordinated management of clients. This is achieved without any modification in the STA which runs standard $802.11$.

In \cite{handoff2} a distributed solution using LVAPs was introduced, proposing a protocol for the direct exchange of information between APs. One limitation of this proposal is that, due to the absence of a central controller, each AP has to build a list of neighbor APs by itself. In \cite{odin1}, a solution based on LVAPs, managed by a central controller, was presented. It combines two Southbound protocols: OpenFlow and \textit{Odin}. OpenFlow tells the internal switches of the APs where to steer the traffic and the \textit{Odin} protocol is in charge of the wireless issues. The central manager of the Wi-Fi network runs within the SDN controller, thus combining both functionalities. The controller is in charge of creating an LVAP for each terminal which consists of a tuple with four fields: the real MAC of the STA, a \textit{fake} MAC for the AP to communicate with the STA, the IP of the STA and the SSID to be used in the communication.

In addition, radio resource management algorithms run as applications on top of the controller, enabling an optimal and dynamic distribution of the STAs between the APs. However, this solution assumes that all the APs operate in the same channel, making it impossible to perform an adequate frequency planning, which constitutes a severe limitation for its use in real deployments. Finally, this solution presents a scalability limitation, as sending broadcast beacons make no sense: each LVAP has to periodically unicast beacons to its station using the corresponding \textit{fake} MAC of the AP.

Fast and seamless handovers are essential part of all these solutions because the STAs can be redistributed dynamically. Therefore, in order not to interrupt the user's session, a seamless AP re-assign can be very convenient when a user is walking, or whenever a load balancing decision is made. In this context, the contribution of the present work is the development and testing of a WLAN centralized solution able to provide seamless handovers between APs in different channels, meeting the QoS requirements of real-time services. In order to develop a good solution, the next conditions have to be accomplished:

\begin{itemize}
\item No changes should be required in the terminal and only low-cost APs can be used.
\item As the users may walk while using real-time services, the handoffs must work correctly at walking speed. Therefore, the solution must meet the quality demands of real-time services with tight latency constraints.
\item The input parameters for the handoff decision should not be limited to the signal level; i.e. the use of other parameters must be possible.
\item APs must be in different channels to reduce the scalability issues caused by the beacon generation for each STA, as reported in \cite{odin1}; so inter-channel handoffs must be possible.
\end{itemize}

All in all, the solution must allow the network to control the mobility and to select the best moment for the handoff. The proposed solution should also be able to separate control and data planes, allowing an abstraction of the underlying technology for applications and services, making it able to interact with other technologies (e.g. 3G or 4G).

\section{Proposed handoff scheme}

The framework presented in \cite{odin1} has been taken as a base to add new functionalities\footnote{All the software components of the solution are available at https://github.com/Wi5}. It should be noted that the fact of using a number of different channels introduces a new degree of complexity. In addition, in order to consider the possibility of using different channels, the CSA (\textit{Channel Switch Announcement}) element in an $802.11$ beacon will be employed (as done in \cite{handoff2}) to make the STAs associated to an AP move to a specific channel.

A number of additional elements have been incorporated in the present proposal:
\begin{itemize}
\item An extra Wi-Fi interface has been added to the APs in order to monitor traffic in other channels, without interrupting the normal operation of the AP. 
\item Different metrics for making the handoff decision can now be used \cite{wifi1}, and they can be evaluated and weighted in the controller, as it has access to all the information. Therefore, the handoff is controlled by the network, and it does not only depend on power measurements on the STA.
\item The controller has a map including the coordinates of the APs, so it does know the channels of APs in the vicinity of a certain one. This information is useful for initiating scans in correct channels whenever the signal of a STA fades (as a consequence of user's movement).
\item The beacon generation has been modified in order to improve the scalability and to give a better user experience during handoffs. 
\end{itemize}

\begin{figure}[!t]
\centering
\includegraphics[width=2.5in]{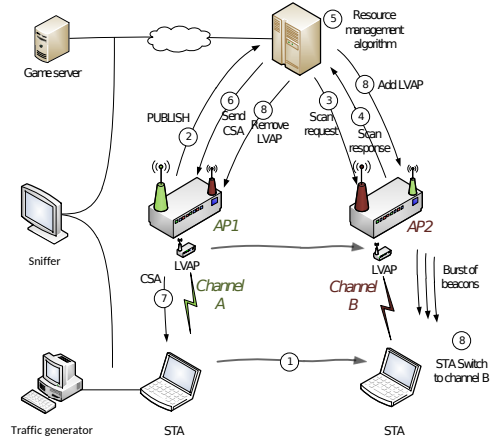}
\vspace{-0.2cm}
\caption{Scheme of an inter-channel reactive handoff using LVAPs.}
\label{figure1}
\end{figure}

Fig. \ref{figure1} shows the proposed handoff scheme, in which we suppose that an STA is associated to AP$1$ (in channel \textit{A}). AP$2$ is in channel \textit{B}. Besides, an Ethernet connection is used to communicate the controller and the APs. The handoff scheme has been designed according to the following process: first, the controller establishes different ``subscriptions" in the APs, in order to raise an event whenever a threshold (noise level, power, etc.) is reached by a STA. When the STA moves $(1)$ away from the Origin AP (AP$1$ in channel \textit{A}), it detects that the signal is below a threshold and sends a \textit{PUBLISH} message to the controller $(2)$. According to its AP map, the controller sends a \textit{Scan Request} message to the neighbor APs $(3)$. For a short period of time, all neighbor APs switch their auxiliary interfaces to channel \textit{A} and listen to packets originated by the STA. If an AP successfully listens to the STA's packets, it sends a \textit{Scan Response} message to the controller $(4)$.

Once the controller has received the \textit{Scan Response} messages from the APs, it runs its algorithms $(5)$ and selects the best suited one for the STA. The decision is: ``\textit{move STA to AP$2$}". Next, the controller tells AP$1$ $(6)$ to send a series of CSAs to the STA $(7)$. They are understood by the STA like a countdown, meaning ``\textit{after N beacons, switch to channel B}".

When the countdown ends, three events must occur in a specific order: $a)$ the STA switches to channel \textit{B} (while AP$1$ does not); $b)$ the controller sends an \textit{Add LVAP} message to AP$2$, which starts sending beacons to the STA in channel \textit{B}, and $c)$ a \textit{Remove LVAP} message is sent to AP$1$. After that moment, the STA starts receiving beacons from AP$2$ in channel \textit{B}. The synchronization of these events is the most critical part of the handoff, as we will see in the subsequent tests and measurements. It should be noted that the STA interprets this as a channel switch carried out by the same AP, so the layer $3$ is not aware of the handoff. Therefore, ongoing communications are only interrupted briefly due to the channel switching. 

Scalability issues may appear when running resource management algorithms, taking into account that the decisions will be made by a central controller. Therefore, the signaling traffic (e.g. monitoring, power control) in the wireless network has to be limited. Something similar will happen in the wired network connecting the APs and the controller, where the signaling traffic should not interfere with the data generated by the clients. In the present case, control and data planes have been separated, so this can be easily achieved. The required processing capacity will also have to be taken into account, as low-cost APs with limited capabilities are being used. This could also be a limitation in the central controller, depending on the number of APs it is managing.


\section{Tests and results}

The test setup (Fig. \ref{figure1}) includes two APs (TP-Link $1043$NDv$2$ with OpenWrt $14.07$) configured in different channels ($4$ and $9$, in the $2.4 \; GHz$ band); the controller runs Debian $8$ (Linux kernel $3.16.0.4$) and the STA runs a Fedora $23$ workstation (Linux kernel $4.4.5.201$). Finally, a DHCP server and a router to access the Internet are included.

\textit{Open vSwitch}\footnote{Open vSwitch, http://openvswitch.org} is installed in the APs, making their internal switch behave as an OpenFlow switch. Floodlight Controller\footnote{Floodlight SDN Controller: http://www.projectfloodlight.org/floodlight} is used, and the resource management algorithms run as applications on top of it. In addition, \textit{Click Modular Router} \cite{click1}, with a specific module, is run in the APs, allowing the interaction with the controller to directly manage the traffic.

The aim of the tests is to analyze the effect of the handoff in the quality of real-time services, in terms of delay and packet loss. As an example of a service with tight latency constraints, we have selected client-to-server traffic of an online game (\textit{Quake} $3$, a popular First Person Shooter using UDP), generated by D-ITG \cite{ditg1}.

In order to avoid common issues while capturing in the radio channel (mainly the ``missing-rate" problem, see \cite{handoff1}), we have included a sniffer (see Fig. \ref{figure1}), so we can accurately obtain the transmitted and received traces by measuring in both (wired) ends of the communication \cite{capturing1}. In addition, the traffic is not generated by the STA itself, but by another machine which sends it to the STA's Ethernet interface, which forwards it to its wireless interface. This setup avoids any degradation of the performance of the applications and allows us to test different Wi-Fi devices with minimal changes.

We have chosen a real environment instead of a free interference scenario. The tests were run in a lab inside a university building, which constitutes a harsh environment with about $15$ APs producing interference. Three different wireless cards have been used (Linksys WUSB$54$GC, WiPi WLAN USB b/g/n, and TP-LINK TL-WN$722$N).

Our initial tests showed the following behavior: when the STA switches its channel, it remains idle for a period. The duration of this period depends on hardware characteristics, the driver and the network implementation. After receiving a certain number of beacons in the new channel, the STA continues the sending of packets. For this reason, inter-beacon time has to be kept low in order to obtain a fast and seamless handoff. 

A trade-off appears: on the one hand, a smaller inter-beacon time permits a faster handoff. On the other hand, high beacon rates can negatively impact the network performance, since broadcast beacons cannot be used, as explained before.

The solution we propose to solve this trade-off consists of defining two different beacon rates: a \textit{low frequency} one, to be used when the STA remains in the same AP; and a \textit{high frequency} one, used for sending a burst of beacons during the handoff. Therefore, when the controller makes a handoff decision, it instructs the destination AP to send a burst of beacons using the high rate. Thus, the wireless card of the STA will hear the required number of beacons in the new channel, and continue its normal operation.

The values of these frequencies have to be selected properly. In the case of the low frequency, we have followed vendor's recommendations (between $50$ and $100 \; ms$). For the \textit{high frequency}, tests with different inter-beacon times, namely $5, 10, 20, 30, 40$ and $50 \; ms$ have been performed, in order to observe the effect of this parameter and to adjust the beacon frequency during the handoff. 

The controller has been configured to force a handoff to the STA every $30 \; s$; besides, every test include $20$ handoffs. The results are reported in terms of packet loss and delay. The packet loss rate can be easily obtained because all transmitted and received traces are known. We can differentiate two causes of packet loss: some packets are lost during the handoff and others are randomly lost due to wireless issues such as interference, packet injection errors and others. The sniffer also allows us to obtain the delay for each packet.

We estimate the handoff time, from the application's point of view, as follows: when the STA starts the channel switch, the wireless interface drops some UDP packets during certain period while switching, and then a number of packets are dropped in a burst. Therefore, we calculate the handoff time as the gap between the last transmitted packet in the \textit{transmitted trace} and the first received packet in the \textit{received trace}. 

\begin{figure*}[!t]
\centering
\includegraphics[width=7in,height=3in]{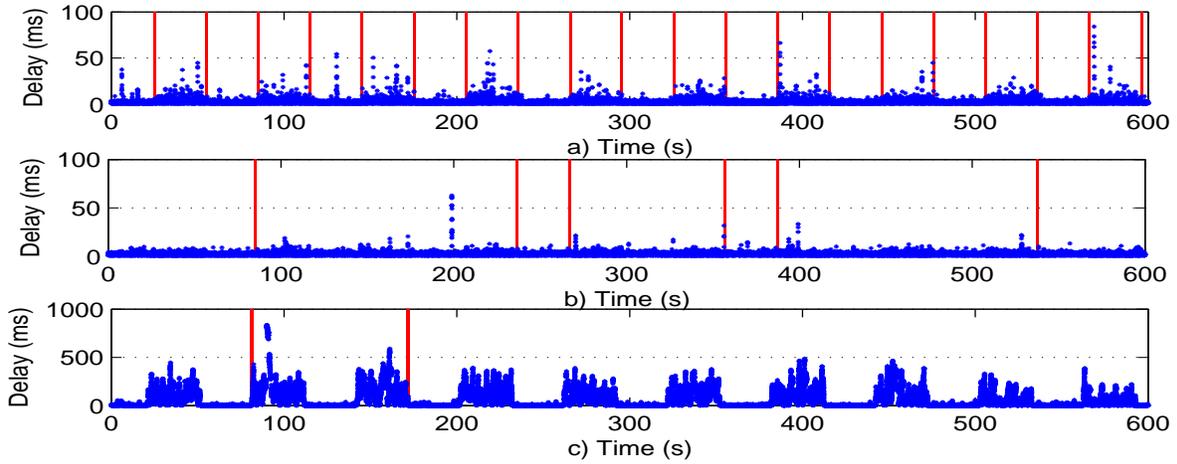}
\vspace{-0.5cm}
\caption{Packet loss (total and caused by the handoff) comparison for different Wi-Fi devices.}
\label{figure2}
\end{figure*}

Fig. \ref{figure2} shows the obtained delay for each packet for a $600 \; s$ transmission when the frequency of the burst of beacons is set to $10 \; ms$. The results are presented for the three wireless cards mentioned above. Vertical lines have been included in order to show the moments in which the handoffs occur. In Fig. \ref{figure2}$a$, we can identify all the $20$ handoffs, but in Fig. \ref{figure2}$b$ and \ref{figure2}$c$, there are cases in which handoffs are undetectable. A handoff cannot be detected in these cases: $a)$ no packets are lost, or $b)$ we cannot distinguish between packets lost by the handoff or by the wireless channel. These undetectable handoffs represent a switch between APs with very good quality. On the other hand, we can observe that the handoff does not increase packet delay; i.e. peaks of delay are not correlated with the moments where handoffs happen.

In Table \ref{table1}, the total packet loss rate and the loss rate caused by the handoff are shown, using different values for the frequency of the burst of beacons. The devices behave in different ways, depending on the beacon frequency: for the Linksys and WiPi devices we can observe a very low level of packet loss attributable to the handoff, especially when the inter-beacon time remains in the lowest levels. However, in the case of TP-Link, all the losses can be attributed to the handoff, and random loss is zero.

\begin{table}[!t]
\renewcommand{\arraystretch}{1.0}
\caption{Packet loss (total and caused by the handoff) comparison for different Wi-Fi devices.}
\label{table1}
\vspace{-0.2cm}
\centering
\scalebox{0.9}[1]{
	\begin{tabular}{c|cc|cc|cc}
		\toprule
		\multirow{2}{*}{Burst} & \multicolumn{6}{c}{Packet loss $ (\%) $} \\
		\cline{2-7}
		 & \multicolumn{2}{c}{Linksys} & \multicolumn{2}{c}{WiPi} & \multicolumn{2}{c}{TP-Link} \\
		 \cline{2-7}
		$ (ms) $ & Total & Handoff & Total & Handoff & Total & Handoff \\
		\midrule 
		$ 5 $ & $ 9.10 $ & $ 0.03 $ & $ 1.24 $ & $ 0.01 $ & $ 0.31 $ & $ 0.31 $ \\
		$ 10 $ & $ 5.64 $ & $ 0.00 $ & $ 1.09 $ & $ 0.01 $ & $ 0.33 $ & $ 0.33 $ \\
		$ 20 $ & $ 4.11 $ & $ 0.05 $ & $ 1.39 $ & $ 0.05 $ & $ 0.36 $ & $ 0.36 $ \\
		$ 30 $ & $ 2.61 $ & $ 0.01 $ & $ 0.80 $ & $ 0.07 $ & $ 0.40 $ & $ 0.40 $ \\
		$ 40 $ & $ 1.43 $ & $ 0.05 $ & $ 0.16 $ & $ 0.14 $ & $ 0.48 $ & $ 0.48 $ \\
		$ 50 $ & $ 0.08 $ & $ 0.04 $ & $ 0.30 $ & $ 0.26 $ & $ 0.60 $ & $ 0.60 $ \\
		\bottomrule
	\end{tabular} }
\end{table}

Table \ref{table2} shows the handoff times for each burst inter-beacon time, and the accumulative percentage of the detected handoffs (e.g. for the Linksys device, when the inter-packet time of the burst is set to $10 \; ms$, $90\%$ of the handoffs last less than $73.80\; ms$). Again, we can see different behaviors for each device: while the TP-Link has the longest handoff time, the other devices show significantly lower values.

\begin{table}[!t]
\renewcommand{\arraystretch}{1.0}
\caption{Percentage (Acc) of times that handoff time can reach.}
\label{table2}
\vspace{-0.2cm}
\centering
\scalebox{0.8}[1]{
	\begin{tabular}{c|cc|cc|cc}
		\toprule
		\multirow{2}{*}{Burst} & \multicolumn{6}{c}{Handoff time} \\
		\cline{2-7}
		 & \multicolumn{2}{c}{Linksys} & \multicolumn{2}{c}{WiPi} & \multicolumn{2}{c}{TP-Link} \\
		 \cline{2-7}
		$ (ms) $ & Acc $ (\%) $ & Time $ (ms) $ & Acc $ (\%) $ & Time $ (ms) $ & Acc $ (\%) $ & Time $ (ms) $ \\
		\midrule 
		$ 5 $ & $ 90.00 $ & $ 223.89 $ & $ 80.00 $ & $ 28.55 $ & $ 0.00 $ & $ 89.86 $ \\
		$ 10 $ & $ 90.00 $ & $ 73.80 $ & $ 70.00 $ & $ 22.29 $ & $ 0.00 $ & $ 87.01 $ \\
		$ 20 $ & $ 75.00 $ & $ 53.13 $ & $ 65.00 $ & $ 35.44 $ & $ 0.00 $ & $ 89.98 $ \\
		$ 30 $ & $ 70.00 $ & $ 59.00 $ & $ 50.00 $ & $ 30.96 $ & $ 0.00 $ & $ 92.58 $ \\
		$ 40 $ & $ 35.00 $ & $ 46.26 $ & $ 35.00 $ & $ 43.27 $ & $ 0.00 $ & $ 93.10 $ \\
		$ 50 $ & $ 10.00 $ & $ 66.80 $ & $ 10.00 $ & $ 61.80 $ & $ 0.00 $ & $ 109.10 $ \\
		\bottomrule
	\end{tabular} }
\end{table}

We have observed that the behavior of each wireless device differs in terms of handoff time and packet loss. In some devices, the handoff time is very small, which produces a small packet loss rate, comparable to the loss produced by the normal interference when the STA is in the same AP. In these cases, it is not possible to determine the moment when a handoff occurs, neither how long it lasts. There exist other cases in which no packets are lost, due to the combination of a fast handoff and the optimal radio channel conditions.

\section{Conclusions and future work}

In this work, a SDN enterprise WLAN solution using a central controller has been presented and tested. By means of virtual APs, it is able to provide seamless inter-channel handoffs. We have improved the scalability by the use of a new handoff scheme and a different beacon generation rate during the handoff. Results have been obtained in terms of delay and packet loss for real-time services using three devices of different vendors. The results show that the handoff scheme presented in this paper does not increase packet delay, and packet loss rate is very low (even zero) depending on the wireless devices. As future work, we plan to explore how to introduce support for roaming between Wi-Fi networks and between Wi-Fi and $3/4G$ networks (vertical handoff).


%



%
%

\ifCLASSOPTIONcaptionsoff
  \newpage
\fi



\bibliographystyle{IEEEtran}
\bibliography{references}

\begin{thebibliography}{10}
\providecommand{\url}[1]{#1}
\csname url@samestyle\endcsname
\providecommand{\newblock}{\relax}
\providecommand{\bibinfo}[2]{#2}
\providecommand{\BIBentrySTDinterwordspacing}{\spaceskip=0pt\relax}
\providecommand{\BIBentryALTinterwordstretchfactor}{4}
\providecommand{\BIBentryALTinterwordspacing}{\spaceskip=\fontdimen2\font plus
\BIBentryALTinterwordstretchfactor\fontdimen3\font minus
  \fontdimen4\font\relax}
\providecommand{\BIBforeignlanguage}[2]{{%
\expandafter\ifx\csname l@#1\endcsname\relax
\typeout{** WARNING: IEEEtran.bst: No hyphenation pattern has been}%
\typeout{** loaded for the language `#1'. Using the pattern for}%
\typeout{** the default language instead.}%
\else
\language=\csname l@#1\endcsname
\fi
#2}}
\providecommand{\BIBdecl}{\relax}
\BIBdecl

\bibitem{sdn1}
R.~Riggio, T.~Rasheed, and M.~K. Marina, ``{Poster: programming
  software-defined wireless networks.}'' in \emph{{MobiCom}}, S.-J. Lee,
  A.~Sabharwal, and P.~Sinha, Eds.\hskip 1em plus 0.5em minus 0.4em\relax ACM,
  p. 413–416.

\bibitem{sdn2}
R.~Riggio, T.~M. Rasheed, and R.~Narayanan, ``{Virtual network functions
  orchestration in enterprise WLANs.}'' in \emph{{IM}}, R.~Badonnel, J.~Xiao,
  S.~Ata, F.~D. Turck, V.~Groza, and C.~R.~P. dos Santos, Eds.\hskip 1em plus
  0.5em minus 0.4em\relax IEEE, p. 1220–1225.

\bibitem{handoff1}
A.~Mishra, M.~Shin, and W.~A. Arbaugh, ``{An empirical analysis of the IEEE
  802.11 MAC layer handoff process.}'' \emph{Computer Communication Review},
  no.~2, p. 93–102.

\bibitem{lvap1}
Y.~Grunenberger and F.~Rousseau, ``{Virtual Access Points for Transparent
  Mobility in Wireless LANs.}'' in \emph{{WCNC}}.\hskip 1em plus 0.5em minus
  0.4em\relax IEEE, p. 1–6.

\bibitem{handoff2}
M.~E. Berezin, F.~Rousseau, and A.~Duda, ``{Multichannel Virtual Access Points
  for Seamless Handoffs in IEEE 802.11 Wireless Networks.}'' in \emph{{VTC
  Spring}}.\hskip 1em plus 0.5em minus 0.4em\relax IEEE, p. 1–5.

\bibitem{odin1}
J.~Schulz-Zander, L.~Suresh, N.~Sarrar, A.~Feldmann, T.~Hühn, and R.~Merz,
  ``{Programmatic Orchestration of {WiFi} Networks},'' in \emph{{2014 USENIX
  Annual Technical Conference (USENIX ATC 14)}}.\hskip 1em plus 0.5em minus
  0.4em\relax Philadelphia, PA: USENIX Association, Jun., p. 347–358.

\bibitem{wifi1}
V.~Mhatre and K.~Papagiannaki, ``{Using smart triggers for improved user
  performance in 802.11 wireless networks.}'' in \emph{{MobiSys}},
  P.~Gunningberg, L.~Åke Larzon, M.~Satyanarayanan, and N.~Davies, Eds.\hskip
  1em plus 0.5em minus 0.4em\relax ACM, p. 246–259.

\bibitem{click1}
E.~Kohler, R.~Morris, B.~Chen, J.~Jannotti, and M.~F. Kaashoek, ``{The click
  modular router},'' \emph{ACM Trans. Comput. Syst.}, no.~3, p. 263–297, Aug.

\bibitem{ditg1}
A.~Botta, A.~Dainotti, and A.~Pescapè, ``{A tool for the generation of
  realistic network workload for emerging networking scenarios.}''
  \emph{Computer Networks}, no.~15, p. 3531–3547.

\bibitem{capturing1}
L.~Zabala, A.~Ferro, and A.~Pineda, ``{Modelling packet capturing in a traffic
  monitoring system based on Linux},'' in \emph{{Performance Evaluation of
  Computer and Telecommunication Systems (SPECTS), 2012 International Symposium
  on}}, 2012, p. 1–6.

\end{thebibliography}
\end{document}